\documentclass[reprint, superscriptaddress, nofootinbib]{revtex4-1}
\usepackage{graphicx}
\usepackage{color}

\usepackage{amsmath}
\usepackage{amssymb}
\usepackage{amsfonts}

\newcommand{\fsr}{\ensuremath{f_\mathrm{fsr}}}

\DeclareMathOperator{\sinc}{sinc}

\begin{document}

\title{
Frequency-Dependent Responses in $3^\mathrm{rd}$ Generation Gravitational-Wave Detectors
}

\author{Reed Essick}
\affiliation{Department of Physics and Kavli Institute for Astrophysics and Space Research, Massachusetts Institute of Technology, Cambridge, MA 02139, USA}
\affiliation{LIGO Laboratory, Massachusetts Institute of Technology, Cambridge, MA 02139, USA}

\author{Salvatore Vitale}
\affiliation{Department of Physics and Kavli Institute for Astrophysics and Space Research, Massachusetts Institute of Technology, Cambridge, MA 02139, USA}
\affiliation{LIGO Laboratory, Massachusetts Institute of Technology, Cambridge, MA 02139, USA}

\author{Matthew Evans}
\affiliation{Department of Physics and Kavli Institute for Astrophysics and Space Research, Massachusetts Institute of Technology, Cambridge, MA 02139, USA}
\affiliation{LIGO Laboratory, Massachusetts Institute of Technology, Cambridge, MA 02139, USA}

\begin{abstract}
Interferometric gravitational wave detectors are dynamic instruments.
Changing gravitational-wave strains influence the trajectories of null geodesics and therefore modify the interferometric response.
These effects will be important when the associated frequencies are comparable to the round-trip light travel time down the detector arms.
The arms of advanced detectors currently in operation are short enough that the strain can be approximated as static, but planned 3$^\mathrm{rd}$ generation detectors, with arms an order of magnitude longer, will need to account for these effects.
We investigate the impact of neglecting the frequency-dependent detector response for compact binary coalescences and show that it can introduce large systematic biases in localization, larger than the statistical uncertainty for 1.4-1.4$M_\odot$ neutron star coalescences at $z\lesssim1.7$.
Analysis of $3^\mathrm{rd}$ generation detectors therefore must account for these effects.
\end{abstract}

\maketitle

\section{Introduction}
\label{section: introduction}

Gravitational-wave (GW) detectors, such as advanced LIGO~\cite{LIGO} and Virgo~\cite{Virgo}, have already enjoyed great success~\cite{GW150914, GW151226, GW170104, O1BBH} and will continue to expand our knowledge of the universe in the coming years~\cite{ObservingScenarios}.
However, planning for $3^\mathrm{rd}$ generation detectors, such as Cosmic Explorer (CE; \cite{CosmicExplorer}) and the Einstein Telescope (ET; \cite{EinsteinTelescope}) has already begun.
Among other technological improvements, $3^\mathrm{rd}$ detectors will have longer arms than the current detectors, with ET proposing a 10 km triangular design~\cite{Freise2009} and CE a 40 km L-shaped detector, an order of magnitude longer than the current LIGO detectors.
Along with the improved sensitivity, the long arms will increase the light travel-time and render dynamical interferometric responses critical within the sensitive band.
These will be important for all signals, regardless of their duration or origin.\footnote{Binary neutron star coalescences may spend several hours in $3^\mathrm{rd}$ generation detectors' sensitive band and therefore the Earth's rotation may also be important. We neglect the Earth's rotation in order to focus on the impact of dynamic interferometric responses alone.}

Several discussions of interferometric GW detectors' frequency dependence already exist in the literature.
Typically, these studies have focused on the response's impact for continuous wave or extremely high-frequency signals (see, e.g.,~\cite{Christensen1990, Christensen1992, malik2008, malik2009}) and have focused on either initial or advanced ground-based detectors~\cite{siggNote, malikNote} or space-based interferometers (see, e.g.,~\cite{Schilling1997, Rubbo2004}).
We instead consider compact binary coalescences (CBCs) detectable by 3$^\mathrm{rd}$ generation ground-based detectors, particularly non-spinning binary neutron star coalescences containing canonical 1.4-1.4$M_\odot$ components.
CBC's signal-to-noise ratios will be dominated by the low-frequency parts of the waveform, although they may also contain high frequency support.
It is often claimed that localization is dependent on the high-frequency signal, but we show that neglecting the frequency dependence of detectors' responses can severely bias localization regardless of signal morphology or duration, including for signals with only low-frequency support that last for $\ll$1 sec.

We begin by describing the basic physical mechanism behind interferometric GW detection in \S\ref{section: basics}, including the impact of frequency dependence on the detector sensitivity in \S\ref{section: antenna response}.
Implications for source localization are discussed in \S\ref{section: implications}, and we conclude in \S\ref{section: conclusions}.
While there may be many more issues associated with neglecting the frequency dependence of our detectors' responses than we discuss here (see~\cite{VitaleInPrep}), the frequency dependence of detector responses must be correctly incorporated into planning for 3$^\mathrm{rd}$ generation detectors.

\section{Basics of the measurement}
\label{section: basics}

In one way or another, interferometric GW detectors operate by timing light's round-trip down their arms and back.
By recording differences in the travel time for multiple arms, they are able to reject many sources of noise, often called common-mode noise, and obtain higher sensitivities than a single arm alone.
Nevertheless, a single arm is sensitive to GWs, and the individual response of each arm can separately affect the recorded signal.

Let us begin by considering the static limit,\footnote{This is sometimes called the \textit{long-wavelength approximation}; the associated GW wavelengths are much longer than the detector's arms.}in which the GW is described by a constant spatial metric perturbation $h_{ij}$.
Throughout this work, latin indices ($i$, $j$, etc) run over spatial dimensions only.
The change in length of a single arm, oriented parallel to the unit-vector $e_i$, is given by $\delta L = L (h_{ij}/2) e^i e^j$, and the factor of $1/2$ comes from expanding the perturbed metric and keeping only linear terms in the perturbation.
By measuring the difference between two arms, we expect our interferometric readout to be
\begin{align}
  \delta V & = \frac{\delta L_x - \delta L_y}{L} \nonumber \\
           & = \frac{1}{2}\left( e^i_x e^j_x - e^i_y e^j_y \right) h_{ij} \label{equation: dV}
\end{align}
which naturally leads the definition of the detector tensor $D^{ij} \equiv (e^i_x e^j_x - e^i_y e^j_y)/2$ and the antenna response for each polarization
\begin{equation}\label{equation: antenna response}
    F_{+,\times} = D_{ij} \varepsilon_{+,\times}^{ij}
\end{equation}
where $\varepsilon_{+,\times}^{ij}$ is the polarization tensor for the $+$ and $\times$ polarizations, respectively~\cite{Anderson2001}.\footnote{$F_{+,\times}$ is often written in terms of three angles: the spherical coordinates $\theta$ and $\phi$ along with a polarization angle $\psi$, which is equivalent to the rotation between the wave-frame's x- and y-axes and the detector's x- and y-axes when $\theta=0$ (directly overhead).}
In this limit, the antenna response is purely a projection effect that maps the strains in the wave-frame onto the detector.
In reality, the dynamics within the detector add additional dependence on both the GW frequency and the direction of propagation relative to the arms.

\begin{figure}
    \begin{center}
        \includegraphics[width=1.0\columnwidth]{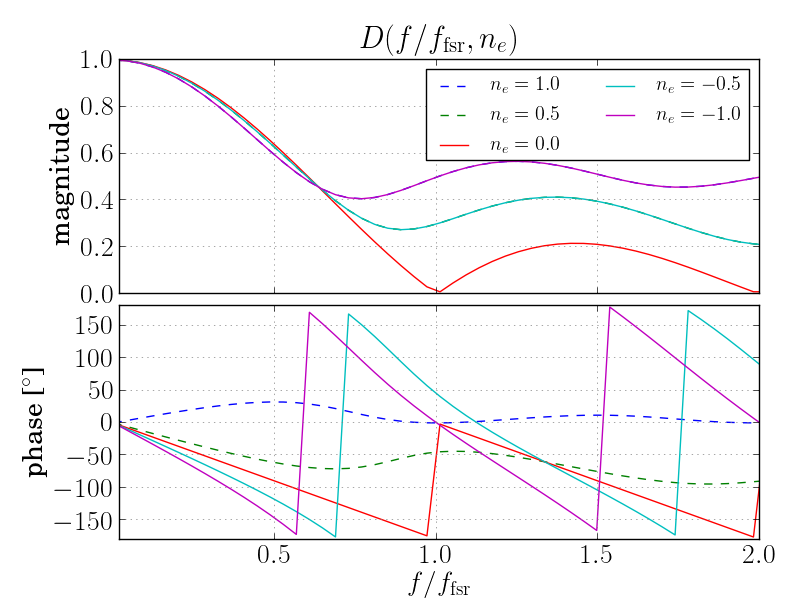}
    \end{center}
    \caption{
        Response of a single arm as a function of frequency for several $n_e$.
        Note that $|D(f,n_e)|=|D(f,-n_e)|$ and the phase is linear in the frequency for many, but not all, $n_e$.
    }
    \label{figure: D}
\end{figure}

Following the procedure outlined in~\cite{malikNote, malik2008, malik2009}, we analyze the round trip travel time between fixed coordinate positions along null geodesics.
Consider a monochromatic GW plane-wave traveling in the direction defined by $n_i$. 
In this case, the GW strain along the arm is given by 
\begin{equation}
    h(t-n_e x/c) = \left(h_+ \varepsilon_+^{ij} + h_\times \varepsilon_\times^{ij}\right) e^{-i\omega(t-n_e x/c)} e_i e_j
\end{equation}
where $n_e \equiv n_i e^i$.
We can then determine the time taken to travel along null geodesics via $cdt = \pm(1+h/2) dx$, recognizing that $dx/dt > 0$ on the outbound trip and $dx/dt < 0$ on the return.
Again, we refer to~\cite{malikNote, malik2008, malik2009} for a more complete derivation, but the fractional change in the travel time for a single arm in the Fourier domain\footnote{We define $\tilde{x}(f) = \int dt\, e^{-2\pi i f t} x(t)$} is
\begin{multline}
    D(f, n_e) \equiv \frac{c}{8\pi i f L} \left(\frac{1-e^{-2 \pi i f (1-n_e)L/c}}{1-n_e} \right. \\
        \left. - e^{-4\pi i f L/c}\frac{1-e^{+2\pi i f (1+n_e)L/c}}{1+n_e} \right),
\end{multline}
While this form clearly shows the contributions from the outbound (first term) and return (second term) parts of the trip, we find it more convenient to express this as
\begin{multline}\label{equation: D}
    D(f, n_e) = \left. \frac{e^{-2\pi i f L/c}}{2(1-n_e^2)} \right( \sinc(2\pi f L/c) - n_e^2\sinc(2\pi f n_e L/c) \\
        \left. - \frac{i n_e}{2\pi f L/c} \left( \cos(2\pi f L/c) - \cos(2\pi f n_e L/c)\right)\right)
\end{multline}
where $\sinc(x)\equiv\sin(x) / x$.
This makes explicit the fact that $|D(f, n_e)| = |D(f, -n_e)|$, as expected from time-reversal symmetry.
Fig.~\ref{figure: D} shows the general behavior of $D(f, n_e)$; the relevant frequency scale corresponds to the unperturbed round-trip travel time along the arm, or the free spectral range ($f_\mathrm{fsr}=c/2L$).

Combining the result from multiple arms allows us to extend the definition of $D^{ij}$ to
\begin{equation}
    D^{ij} = D(f, n_k e_x^k) e_x^i e_x^j - D(f, n_l e_y^l) e_y^i e_y^j
\end{equation}
We also see that 
\begin{equation}\label{equation: static limit}
    \lim\limits_{f\rightarrow0} D(f, n_e) = \frac{1}{2} - \frac{\pi i f L}{c}\left(1-\frac{n}{2}\right)
\end{equation}
in agreement with our analysis of a static strain (Eqn.~\ref{equation: dV}).
Importantly, we note that the antenna responses defined in Eqn.~\ref{equation: antenna response} are transfer functions from the astrophysical strain in the wave-frame to the detector readout.
Therefore, they are complex functions dependent on the direction to the source (and therefore the propagation direction) along with the GW frequency.
This means that the antenna pattern relevant for a single source will evolve in time as GWs from the source evolve in frequency.

\subsection{General Behavior of the Antenna Response}
\label{section: antenna response}

\begin{figure*}
    \begin{minipage}{0.49\textwidth}
        \begin{center}
            \includegraphics[width=1.0\textwidth]{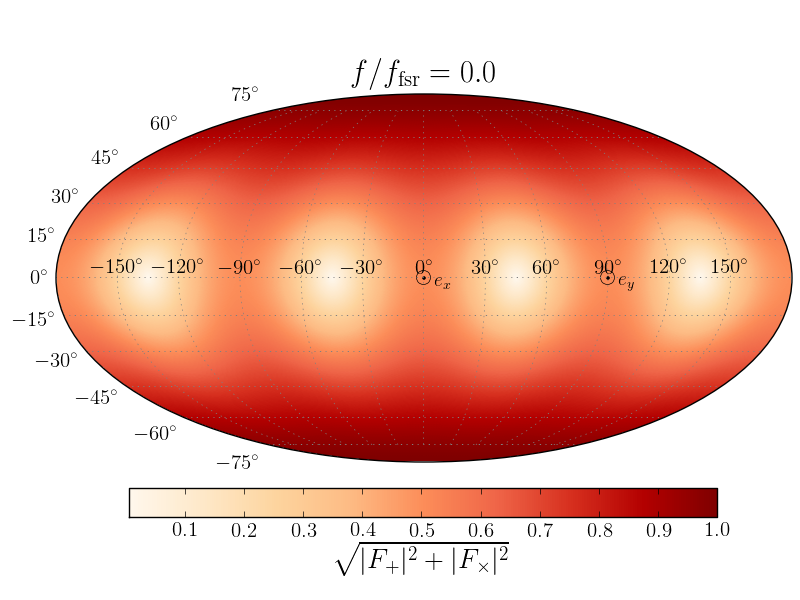} \\
            \includegraphics[width=1.0\textwidth]{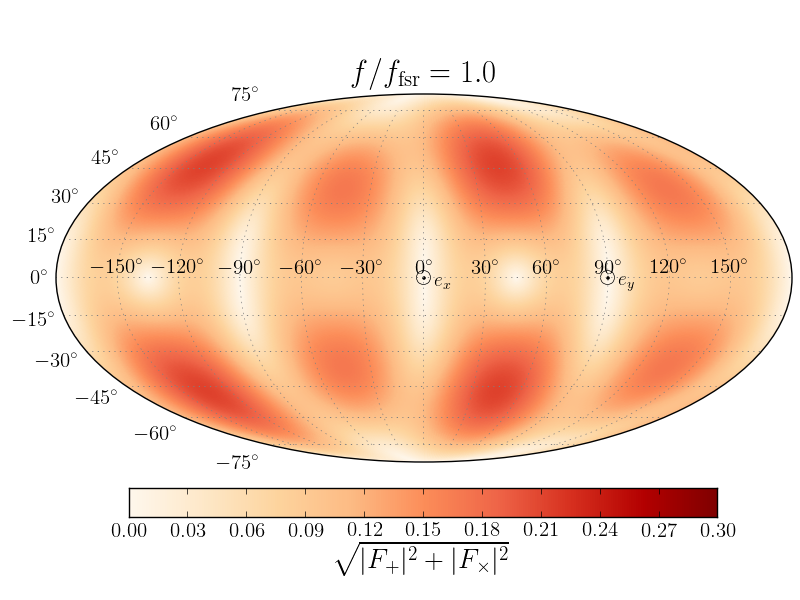}
        \end{center}
    \end{minipage}
    \begin{minipage}{0.49\textwidth}
        \begin{center}
            \includegraphics[width=1.0\textwidth]{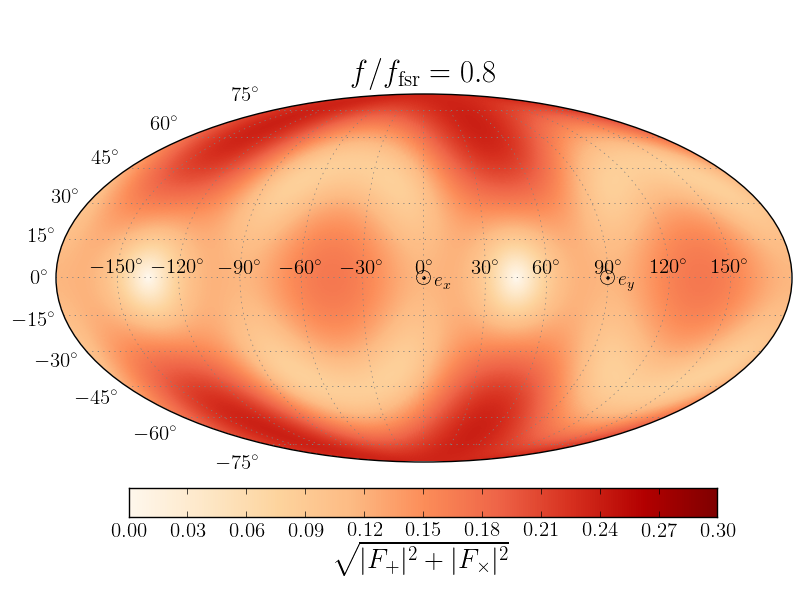} \\
            \includegraphics[width=1.0\textwidth]{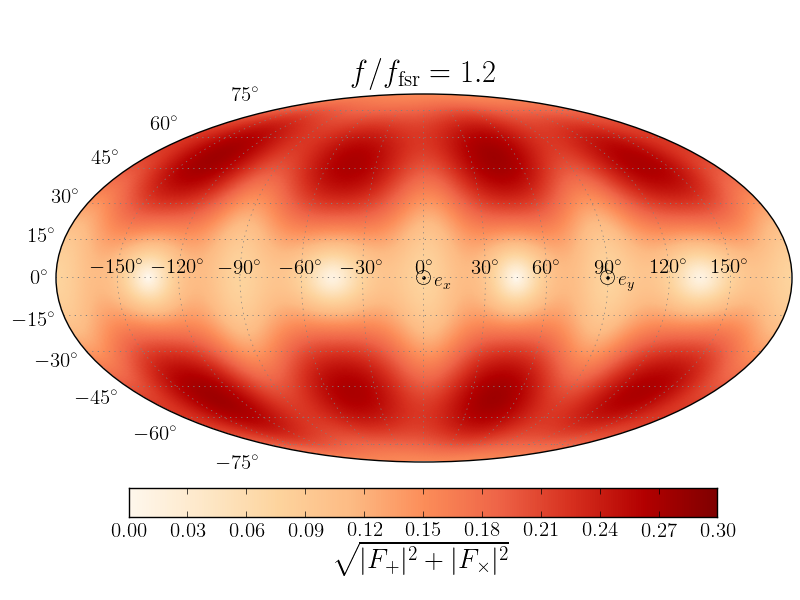}
        \end{center}
    \end{minipage}
    \caption{
        Directional dependence of the overall sensitivity to GWs for a detector aligned with the coordinate axes.
        We note that the color-scale for $f/\fsr=0$ is different than the scale for $f/\fsr>0$ to account for the dramatic change in overall sensitivity (see Fig.~\ref{figure: antenna bode}); the three plots with $f/\fsr>0$ all have the same scale.
    }
    \label{figure: antenna mollweide}
\end{figure*}

\begin{figure*}
    \begin{minipage}{0.49\textwidth}
        \begin{center}
            \includegraphics[width=1.0\textwidth]{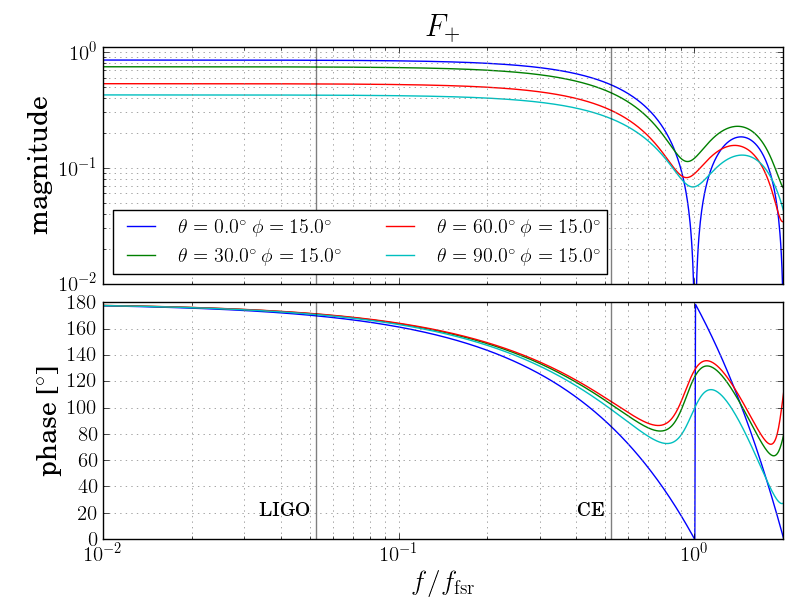}
        \end{center}
    \end{minipage}
    \begin{minipage}{0.49\textwidth}
        \begin{center}
            \includegraphics[width=1.0\textwidth]{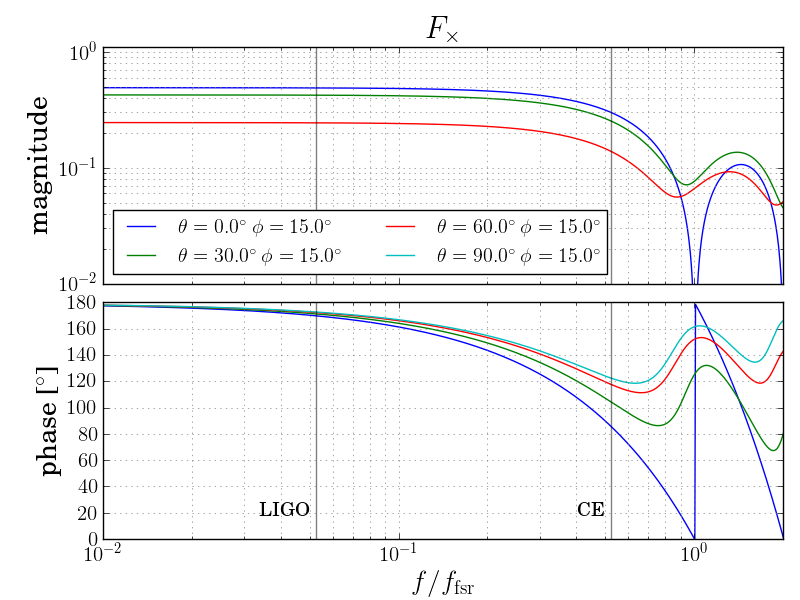}
        \end{center}
    \end{minipage}
    \caption{
        Bode plots of $F_+$ and $F_\times$ at a few source directions ($\theta$, $\phi$) with polarization angle $\psi=0$.
        Note the logarithmically scaled ordinate, in contrast to the linearly scale in Fig.~\ref{figure: D}.
        Grey lines denote the dynamical frequency for neutron stars ($f=(1/2\pi)\sqrt{GM/R^3}\sim1.97\text{ kHz}$) normalized by \fsr~for LIGO and CE, respectively.
        We also note that $F_+<1$ for all frequencies, even at $\theta=0^\circ$, for $\phi=15^\circ$. 
        This is because of a degeneracy between $\phi$ and $\psi$ when $\theta=0^\circ$, which effectively mixes the polarizations.
        Instead, $\sqrt{|F_+|^2+|F_\times|^2}\rightarrow1$ as $f\rightarrow0$ for $\theta=0^\circ$, independent of $\phi$ (see Fig.~\ref{figure: antenna mollweide}).
    }
    \label{figure: antenna bode}
\end{figure*}

The antenna responses can change dramatically as a function of both the source location and the GW frequency.
Fig.~\ref{figure: antenna mollweide} shows the overall sensitivity ($\sqrt{|F_+|^2 + |F_\times|^2}$) as a function of source location at several frequencies for a detector oriented along the coordinate axes ($e_x = (1, 0, 0)$ and $e_y = (0, 1, 0)$).
Similarly, Fig.~\ref{figure: antenna bode} shows both the magnitude and phase of the response to each polarization separately for a few source locations in spherical coordinates defined relative to the detector.

At frequencies small compared to $f_\mathrm{fsr}$, we notice little difference in the overall shape of the antenna response, although the phase does change.
The predominant change is a decrease in the magnitude, which is apparent in the color scales within Fig.~\ref{figure: antenna mollweide} but is more clearly depicted in Fig.~\ref{figure: antenna bode}.
However, when we approach (and exceed) $f_\mathrm{fsr}$, there are large changes to the directional sensitivity.
In particular, the maximum in the detector response directly overhead ($\theta=0$) becomes a zero.
We also note that $\sqrt{|F_+|^2 + |F_\times|^2}$ is only symmetric under rotations of $\pi$ about the $z$-axis for all GW frequencies, which is expected from $l=2$ spherical harmonics.
This is in contrast to the symmetry under rotations of $\pi/2$ in the static limit, which is due to an additional symmetry between $F_+$ and $F_\times$ in that limit.

For a fixed source location, we note the general decrease in $|F_{+,\times}|$ at higher frequencies as well as the significant change in phase, even at $f \lesssim f_\mathrm{fsr}/2$.
The annotations in Fig.~\ref{figure: antenna bode} show the dynamical frequency of a 2$M_\odot$ neutron star with a 12 km radius ($f_\mathrm{ns} = (1/2\pi)\sqrt{GM/R^3}=1.97\, \mathrm{kHz}$) normalized by $f_\mathrm{fsr}$ for LIGO and CE, respectively.
It is clear that LIGO is reasonably approximated by the static limit, but CE can accrue significant phase and a respectable loss in sensitivity at $f\gtrsim 2\, \mathrm{kHz}$.
In particular, we note that the phase introduced is linear in the frequency (Eqn.~\ref{equation: static limit}) for many source locations, which appears as a time offset (Eqn.~\ref{equation: time-offset}), and, given the scale, it is not unreasonable to ask whether neglecting this phase can significantly impact GW measurements.

\section{Implications for Localization}
\label{section: implications}

We focus on the ``bread and butter'' CBCs expected to be observed in great numbers with current and 3$^\mathrm{rd}$ generation detectors (see, e.g., ~\cite{ObservingScenarios, GW170104, Sathyaprakash2010, Regimbau2017, Vitale2017}).
CBCs are dominated by the low frequency parts of their signal, generally where we would expect the static limit to be most appropriate.
These results should apply to all signals as long as $f \ll f_\mathrm{fsr}$, independent of the actual waveform.
We also only focus on source localization instead of enumerating all the possible ways the static limit may be inappropriate for 3$^\mathrm{rd}$ generation detectors.
The point being that as long as there is at least one major shortcoming, the static limit cannot be assumed.

We first note that, because $|F_{+,\times}|$ is approximately constant for $f\ll f_\mathrm{fsr}$, we do not expect to lose much signal-to-noise ratio by neglecting frequency-dependent effects.
Instead, we find that it is the phase that introduces potentially large biases into the reconstructed location of sources, which is mostly due to confusion between the neglected phase and the signal's arrival time at each detector.
To understand this, we first consider the effect of a time-delay on the Fourier transform of the signal
\begin{align}
    \int dt\, e^{-2\pi i f t} h(t+\delta t) & = \int d\tau\, e^{-2\pi i f (\tau - \delta t)} h(\tau) \nonumber \\
                                            & = e^{2\pi i f \delta t} \tilde{h}(f) \label{equation: time-offset}
\end{align}
We note that the additional phase introduced by the time-delay is $\Phi = 2\pi f \delta t$, and therefore any phase that is linear in $f$ could be confused for an analogous time-delay.
Now, the actual phase introduced by the frequency dependence of the antenna response depends on the GW's propagation direction relative to each of the detector's arms, and therefore can easily be different for different detectors.
That implies that the inferred time-delay will differ in each detector, causing a net change in the time-of-arrival difference between the detectors.
Ground-based GW detectors primarily localize signals via triangulation, and therefore these effects, if neglected, introduce biases in the reconstructed source location.

\begin{figure}
    \begin{center}
        \includegraphics[width=1.0\columnwidth]{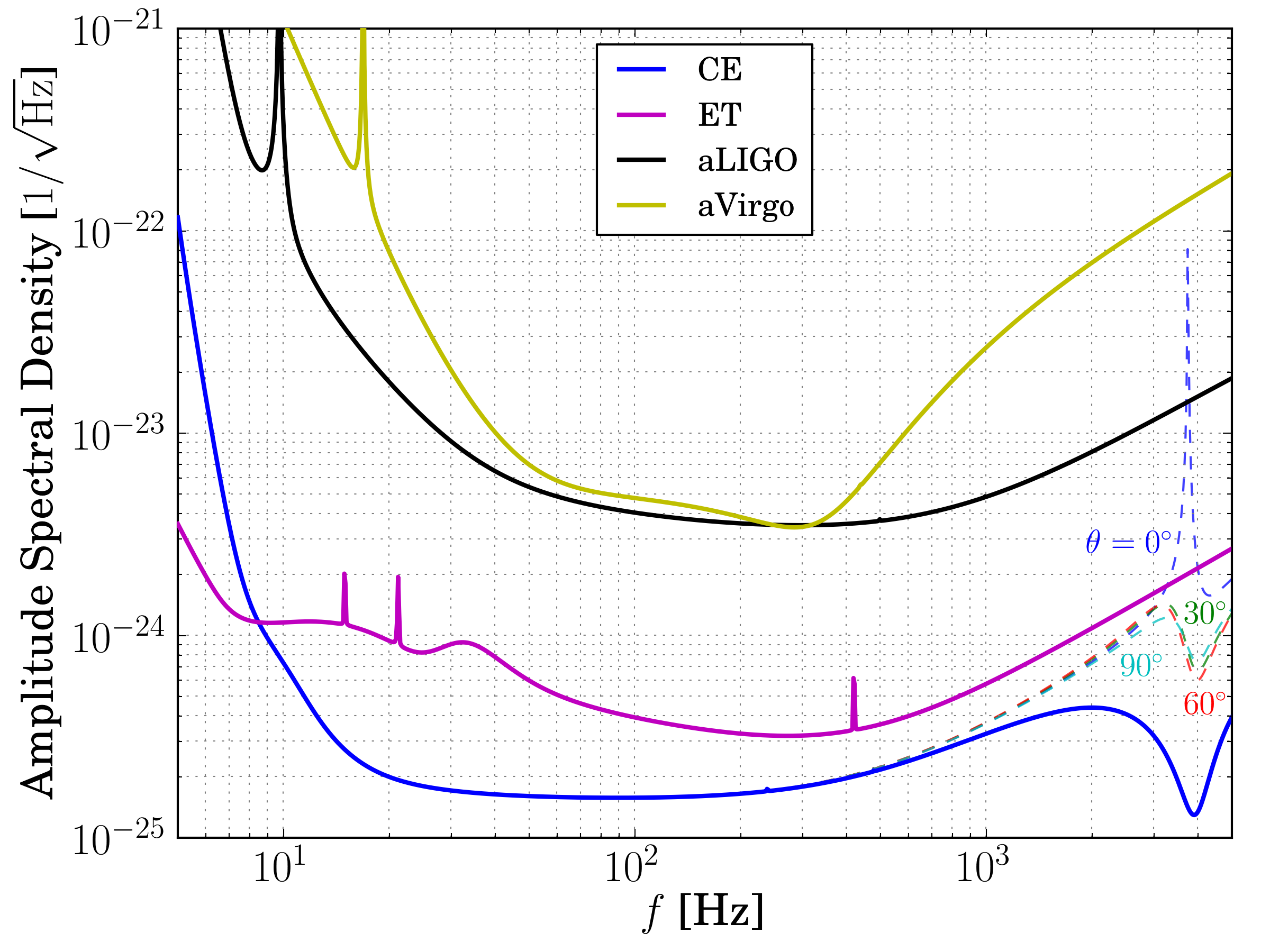}
    \end{center}
    \caption{
        Amplitude Spectral Densities assumed in our simulations; our Monte Carlo regressions compute likelihoods using frequencies $\geq10\text{ Hz}$.
        Because the actual detector response, and therefore the effective noise floor, depends on both the GW frequency and the source location, we show the underlying noise curves neglecting directional dependence in thick solid lines.
        Thin dashed lines demonstrate the variability with source location (normalized by the static limit responses), and colors correspond to the relative orientations in Fig.~\ref{figure: antenna bode}.
        Curves presented in~\cite{CosmicExplorer} assume sources nearly, but not exactly, overhead the detectors.
        We only show the directional dependence for CE because all other detectors are short enough that the effects are small.
    }
    \label{figure: PSDs}
\end{figure}

\begin{figure*}
    \begin{center}
        \includegraphics[width=0.75\textwidth, clip=True, trim=0.40cm 0.75cm 0.20cm 1.75cm]{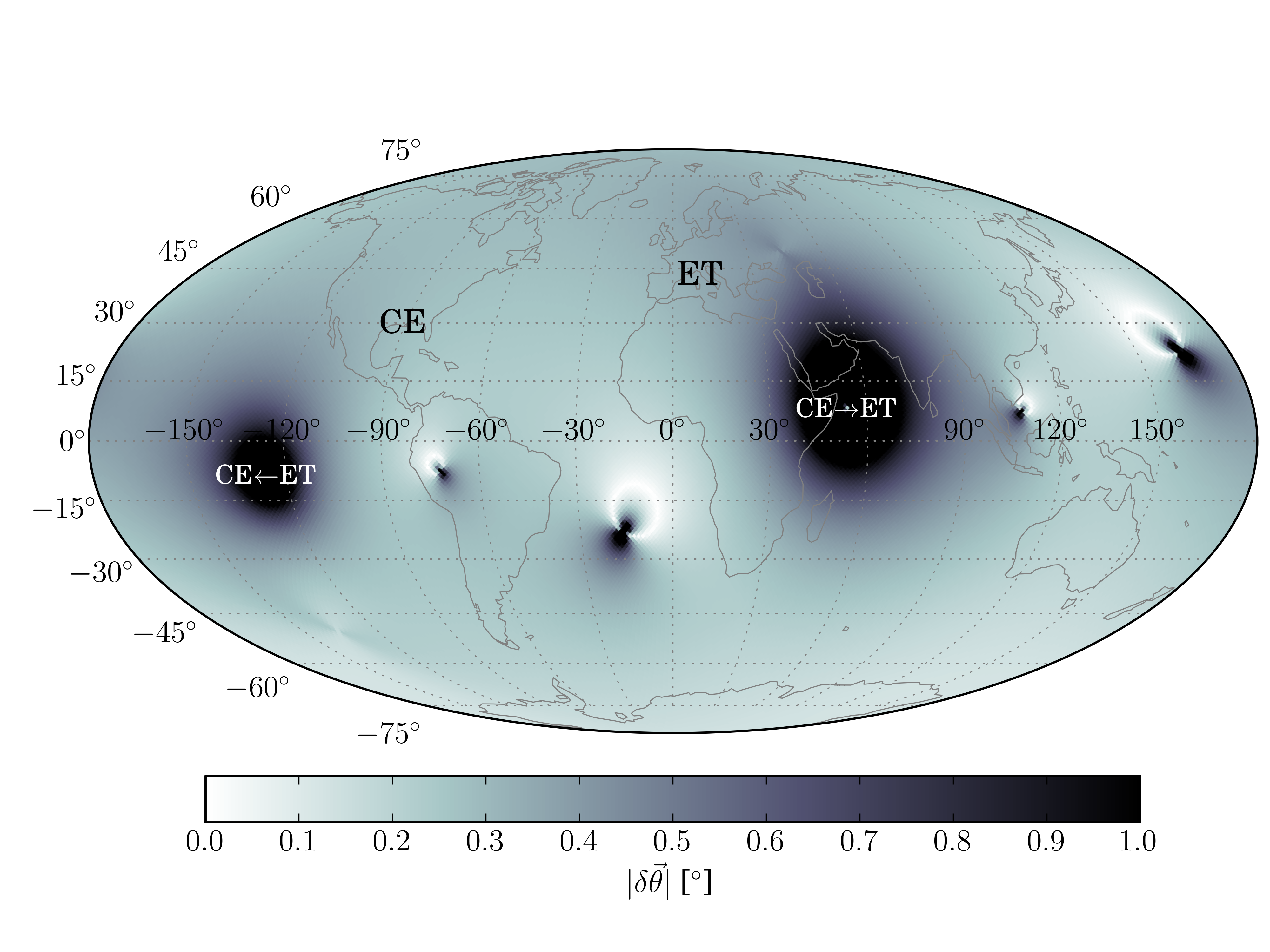} \\
        (a)
    \end{center}
    \begin{minipage}{0.30\textwidth}
        \begin{center}
            \includegraphics[width=1.0\textwidth]{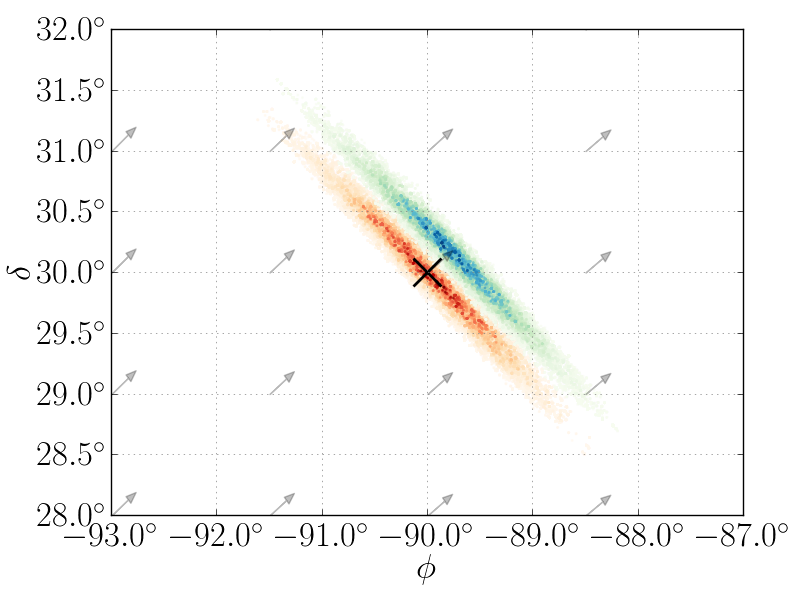} \\
            (b)
        \end{center}
    \end{minipage}
    \begin{minipage}{0.30\textwidth}
        \begin{center}
            \includegraphics[width=1.0\textwidth]{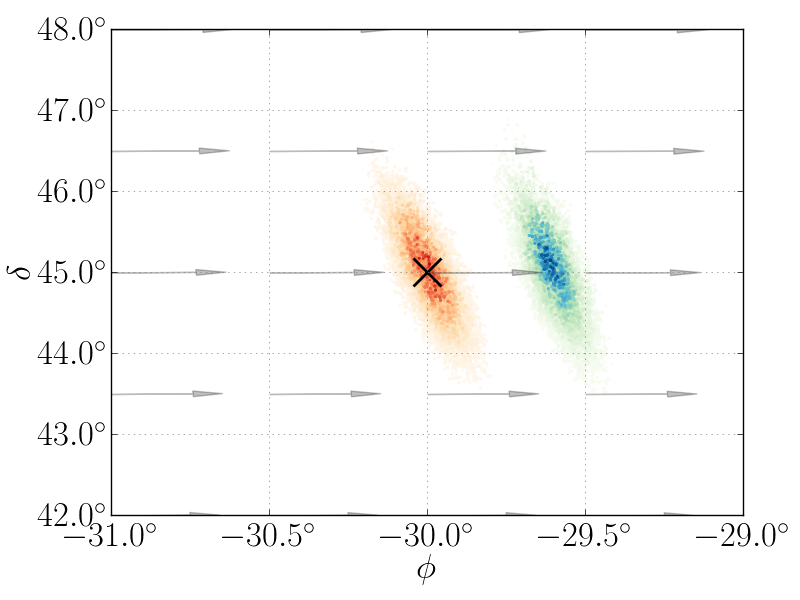} \\
            (c)
        \end{center}
    \end{minipage}
    \begin{minipage}{0.30\textwidth}
        \begin{center}
            \includegraphics[width=1.0\textwidth]{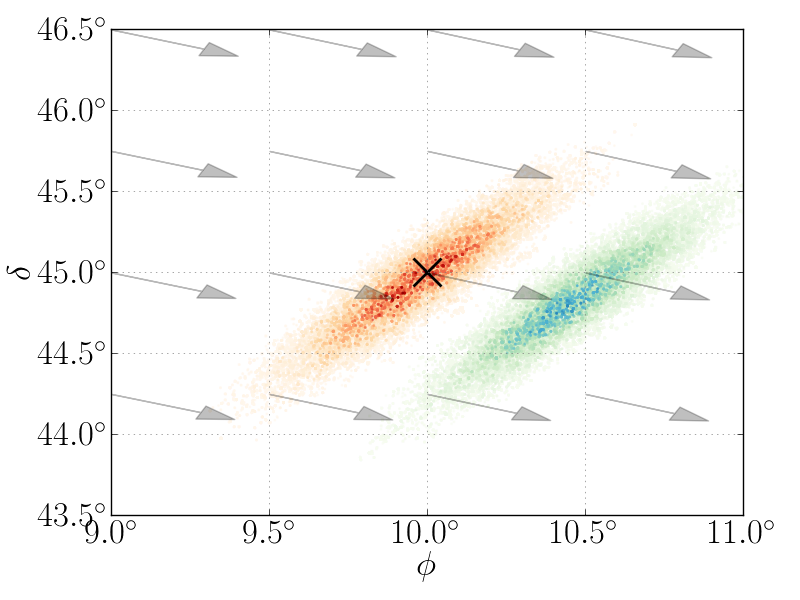} \\
            (d)
        \end{center}
    \end{minipage}
    \caption{
        (a) Predicted biases introduced by neglecting the detector responses' phase for a network with CE located at the current LLO site and ET located at Virgo.
        Large biases are due to single detector response poles and the coordinate singularity between time-delays and triangulation's polar angle. 
        (b, c, d) Monte Carlo estimates with (red) the correct antenna response and (blue) the static limit, along with (arrows) analytic predictions for the bias based on Eqn.~\ref{equation: analytic}.
        The source's true location is shown with a black $\times$.
        Our Monte Carlo simulations only marginalized over extrinsic parameters and neglected cosmological effects (which is a good approximation for $D\sim500\text{ Mpc}$).
        Note that the analytically predicted bias is a good approximation for the full numerical result.
    }
    \label{figure: localization bias}
\end{figure*}

Fig.~\ref{figure: localization bias} demonstrates an analytic approximation for the bias introduced within a network of one CE located and oriented identically to the current LIGO Livingston detector and one vertex of the proposed ET, located at the current Virgo site.
Specifically, we compute the complex phase of a linear combination of the antenna responses for each polarization
\begin{equation}\label{equation: analytic}
    \Phi = \mathrm{arg}\left\{ F_+ \frac{1}{2}(1+\cos^2 \theta_{jn}) + F_\times \cos \theta_{jn} \right\}
\end{equation}
where $\theta_{jn}$ is the inclination angle between the orbit and the wave's propagation direction.
We compute $\Phi$ as a function of frequency and extract the corresponding time-delay through a linear fit.
The difference in this time-delay between detectors is converted to an angular bias through triangulation.
The time-delay introduced by the detector response makes it appear as if the signal was recorded later at CE, and when this effect is neglected it biases the reconstructed location away from CE and toward ET.
CE dominates the effect because of its longer arms.

We note that this crude prediction does not involve the GW morphology in any way, but nonetheless accounts for the vast majority of the bias observed in Monte Carlo regressions\footnote{Our implementation is publicly available~\cite{freqDepAnt} and is based off~\cite{emcee}.} simulating a 1.4-1.4$M_\odot$ binary neutron star coalescence (red and blue posteriors in Fig.~\ref{figure: localization bias}) using the noise curves shown in Fig.~\ref{figure: PSDs}.
What's more, the systematic bias can be larger than the statistical uncertainty in the localization.

An immediate question is whether the bias introduced by assuming the static limit could impact the current detections~\cite{GW150914, GW151226, GW170104, O1BBH}.
Based on Eqn.~\ref{equation: D}, we expect the largest time-delay introduced to be $|\delta t|\lesssim L/c\sim1.33\cdot10^{-5}\text{ sec}$, which corresponds to $\sim0.076^\circ$ for the two LIGO detectors, much less than the statistical uncertainty in typical localization estimates ($\gtrsim 3^\circ$~\cite{GW150914, GW151226, GW170104, LOSC}).
In reality, the bias is likely to be even smaller because of the near alignment and identical arm-lengths of the LIGO detectors.
Heterogeneous networks, like the CE+ET network depicted in Fig.~\ref{figure: localization bias}, often produce larger biases than homogeneous networks because there is less cancellation of the effect between detectors.

Networks involving 40-km scale detectors could produce time-delays an order of magnitude larger ($L/c\sim1.33\cdot10^{-4}\text{ sec}$), and the systematic bias could be $\gtrsim 0.5^\circ$.
For nearby sources, like those of interest for electromagnetic follow-up ($D \lesssim 500\text{ Mpc}$).
This can be much larger than the statistical uncertainty in the localization.
Following the procedure outlined in~\cite{Fairhurst2009, Fairhurst2009Errata}, and including cosmological effects~\cite{Planck}, we find the expected standard deviation in the time-of-arrival difference between two CE detectors is comparable to the systematic bias at $z\sim1.7$ ($D\sim13.1\text{ Gpc}$).
At this distance, a 1.4-1.4$M_\odot$ binary neutron star coalescence would have a single-detector signal-to-noise ratio of $\rho\sim26.5$ and would be easily detected~\cite{Vitale2016}.

We also note that the changing directional dependence of $D(f,n_e)$ could possibly improve localization estimates.
This is not the case for CBCs, which are dominated by $f\ll f_\mathrm{fsr}$.
However, the impact may be larger for core-collapse supernova waveforms, which have more energy at higher frequencies.
The Earth's rotation may also improve localization for long-duration signals (see, e.g., \cite{Schilling1997}), but quantifying this effect for ground-based detectors is beyond the scope of this paper.

\section{Conclusions}
\label{section: conclusions}

Interferometric GW detectors are dynamic instruments that respond differently at different frequencies.
Fundamentally, this is associated with the change in projected strain as the light travels down the arms and back.
Therefore, the free spectral range sets the scale for frequency dependent effects, and $f_\mathrm{fsr}$ can be comparable to the sensitive band for $3^\mathrm{rd}$ generation detectors.

We consider the impact of neglecting out detectors' frequency dependence for canonical 1.4-1.4$M_\odot$ binary neutron star coalescences and find significant biases, even for sources at cosmological distances.
We demonstrate that the bias is due to the neglected phase of the antenna response, which resembles a time-delay and confuses triangulation.
This means the biases will impact all sources, regardless of their waveform morphology, even when $f\ll f_\mathrm{fsr}$.
The bias should be much smaller than statistical uncertainties for current detectors (e.g. LIGO and Virgo), but $3^\mathrm{rd}$ generation detectors, with arms an order of magnitude longer, will be biased beyond statistical uncertainties for $z\lesssim1.7$.

The biases on localization alone show that we must account for interferometric frequency dependence for $3^\mathrm{rd}$ generation detectors for an appreciable fraction of all observable sources in the universe~\cite{Vitale2016}.
Furthermore, the exact detector response is known, is relatively straightforward to implement numerically, and therefore should be incorporated into all planning for and analysis of $3^\mathrm{rd}$ generation detector science.


The authors thank Aaron Buikema for useful discussion and moral support.
The authors also acknowledge the support of the National Science Foundation and the LIGO Laboratory.
LIGO was constructed by the California Institute of Technology and Massachusetts Institute of Technology with funding from the National Science Foundation and operates under cooperative agreement PHY-0.497058.


\bibliography{refs}

\end{document}